\documentclass[preprint,review,12pt]{elsarticle}

\usepackage{graphicx} 
\usepackage{amssymb} 
\usepackage{amsmath} 
\usepackage{lineno}
\usepackage{color,soul}

\usepackage{natbib}
\setcitestyle{authoryear,open={(},close={)}}

\begin{document}
\begin{frontmatter}

\title{Gaussian Markov Random Fields versus Linear Mixed Models for satellite-based PM\textsubscript{2.5} assessment: Evidence from the Northeastern USA}


\author[a]{Ron Sarafian}
\address[a]{Department of Industrial Engineering, Ben Gurion University of the Negev, Be'er Sheva, Israel}
\ead{ronsar@post.bgu.ac.il}

\author[b]{Itai Kloog}
\address[b]{Department of Geography and Environmental Development, Ben Gurion University of the Negev, Be'er Sheva, Israel}

\author[c]{Allan C. Just}
\address[c]{Department of Environmental Medicine and Public Health, Icahn School of Medicine at Mount Sinai, New York, NY, USA}

\author[a]{Johnathan D. Rosenblatt}
\ead{johnros@bgu.ac.il}

\begin{abstract}
Studying the effects of air-pollution on health is a key area in environmental epidemiology. An accurate estimation of air-pollution effects requires spatio-temporally resolved datasets of air-pollution, especially, Fine \emph{Particulate Matter} (PM). Satellite-based technology has greatly enhanced the ability to provide PM assessments in locations where direct measurement is impossible.

Indirect PM measurement is a statistical prediction problem. The spatio-temporal statistical literature offer various predictive models: Gaussian Random Fields (GRF) and Linear Mixed Models (LMM), in particular. GRF emphasize the spatio-temporal structure in the data, but are computationally demanding to fit. LMMs are computationally easier to fit, but require some tampering to deal with space and time.

Recent advances in the spatio-temporal statistical literature propose to alleviate the computation burden of GRFs by approximating them with Gaussian Markov Random Fields (GMRFs). Since LMMs and GMRFs are both computationally feasible, the question arises: which is statistically better? We show that despite the great popularity of LMMs in environmental monitoring and pollution assessment, LMMs are statistically inferior to GMRF for measuring PM in the Northeastern USA. \\
\end{abstract}


\end{frontmatter}


\section{Introduction}
Studying the adverse effects of air-pollution on health is an important topic in environmental epidemiology. The estimation of the effects of air-pollution requires datasets of air-pollutants' concentrations, that can be combined with associated health data. Particulate Matter (PM) is one of the regularly monitored air pollutants, and a major concern in public health \citep{schwartz1996daily,kloog2013long}. PM mass concentrations are measured at ground monitoring stations, usually around urban areas, therefore are limited in terms of spatial coverage. In recent years, remotely sensed satellite data is being used to assess PM. In particular, Aerosol Optical Depth measurements (AOD) have been found to be good predictors of PM. AOD's large temporal and spatial coverage enables assessments of PM in times and locations where direct measurement is impossible \citep{streets2013emissions}. PM assessment from AOD can be seen as a statistical prediction problem, with a spatio-temporal structure.

\emph{Linear Mixed Models} (LMM) have been established as one of the leading methods in environmental air-pollution assessments \citep{chudnovsky2014fine,kloog2015estimating,lee2016spatiotemporal,shtein2018estimating}. LMMs are widely used in environmental studies due to their ability to specifying complex correlation structures, via easy-to-specify random effects. Its popularity in PM prediction stems mainly from the fact that it yields highly accurate predictions at a low computational cost. The main computational burden when fitting large spatio-temporal models is the manipulation of the covariance matrix. LMMs imply a simple sparse block-diagonal covariance structure: a block matrix having main diagonal blocks square matrices, such that the off-diagonal blocks are null matrices. The inverse of LMMs' covariance, i.e. its \emph{Precision matrix}, maintains this sparse structure. Algorithms optimized for sparse matrices require less memory and computing time than algorithms for arbitrary matrices \citep{duff2017direct}.

The spatio-temporal statistical literature typically advocates models with smoothly decaying correlations. The success and popularity of LMMs in air-pollution, with its (non-smooth) ``slab'' structured correlation structure may thus be surprising for someone immersed in the spatio-temporal statistical literature. The most fundamental statistical model for space-time process over continuous domains is the Gaussian Random Field (GRF) \citep{diggle1998model, banerjee2004hierarchical}. A GRF is determined through its mean and covariance functions. Although it has good analytic properties, a GRF is computationally hard to fit. Fitting a GRF with maximum-likelihood, and $N$ data points, involves the inversion of an $N\times N$ covariance matrix, hence, requires $\mathcal{O}(N^3)$ operations. This makes its calculation infeasible for some real-world spatio-temporal PM datasets \citep{porcu2012advances}.

Over the past years, several strategies have been developed to alleviate the computational cost of learning GRFs \citep{heaton2017case}. Sparse precision matrices are at the core of most of these techniques. In a recent line of work, a continuously indexed GRF is approximated by a discretely indexed Gaussian Markov Random Field (GMRF) \citep{lindgren2011explicit, bakka2018spatial}. This approach allows fitting a GRF with a continuously and smoothly decaying covariance function, while computations are performed with the sparse precision matrix of a Markovian process. The learning of the GMRF can be performed in a Bayesian hierarchical framework with Gaussian priors. As such, this model can be described as a Latent Gaussian Model (LGM) \citep{rue2009approximate} that is characterized by a GRF, with a GMRF representation. 

GMRFs, like LMMs, imply sparse precision matrices, so they are efficiently computable. Fitting our GMRF approximation of a GRF with a continuous Mat\'ern covariance takes about $1-1.5$ hours on a $8$-cores machine, for a dataset of size $N=100K$, with $300$ stations and $1,400$ days. Fitting a GMRF is thus more computationally challenging than fitting an LMM, but still feasible. The question now arises: which modeling approach is statistically preferable?

In our work we attempt to compare the statistical accuracy of LMMs to GMRFs, which differ only in their spatial random effects. Our baseline LMM formulation follows several recent studies \citep{kloog2014new, kloog2015estimating}, which use spatial and temporal random effects to model correlations. Their PM predictions are used in a wide range of epidemiological studies \citep[e.g.]{mehta2016long, wang2016estimating, rosa2017prenatal, kloog2018associations}. We emphasize that we are not trying to improve PM predictions in general, but rather, to study the effect of the spatial correlation model. We thus did not try to improve modeling by other means, for example using new predictors, or different temporal modeling.

Our case study is from Northeastern USA. This data includes areas with many dense PM monitoring stations, as well as large areas where stations are scarce. Stations are thus placed on a very non-regular grid. This serves our purpose, since we can study the effect of the correlation model for short-range and long-range predictions, densely-monitored vs.\ sparsely-monitored regions, etc.


\section{Materials and Methods}


\subsection{Study Domain}
The study area contains the Northeastern part of the USA, from Maine to Virginia. It includes urban areas such as New York and Boston, as well as rural areas and largely uninhabited regions. The study period is from January 1st, 2000, to December 31st, 2015, although very often PM data is not available in all stations.


\subsection{PM Monitoring Data}
PM mass concentrations data were obtained from the U.S. Environmental Protection Agency (EPA) Air Quality System (AQS) database. We removed some PM measurements due to lack of reliable AOD observations in certain days, and also excluded days with less than 30 spatial observations and stations with less than 30 temporal observations. The number of PM monitoring stations included is $330$, and the number of time-points (i.e., unique days) is $1439$. The total number of observations in our dataset is $100,791$.


\subsection{AOD Data}
Spectral AOD measures the extinction of the solar beam by particles within an atmospheric column and is a fundamental satellite-derived measurements for PM prediction. The AOD data we use is derived from the Multi-Angle Implementation of Atmospheric Correction (MAIAC) algorithm for Moderate Resolution Imaging Spectroradiometer (MODIS) instrument on the Aqua satellite in $1$km resolution \citep{lyapustin2011multiangle1, lyapustin2011multiangle2}. Valid AOD estimation over heterogeneous landscapes is challenging as it is affected by remote sensing conditions, and measurement errors are difficult to avoid. Therefore, we use a corrected version of the aforementioned AOD product, recently developed by \citet{just2018correcting}. This correction was applied using collocated measures from ground-based AERONET stations through Gradient Boosting mechanism, and has been proven effective in improving MAIAC AOD product over the Northeastern USA without using any of the data we wish to predict at the PM monitoring stations.


\subsection{Non-AOD PM Predictors}
Here also we follow \citet{just2018correcting} in the choice of other PM predictors (see references therein). Meteorologic data was derived from the NCEP North American Regional Reanalysis dataset (daily averages) and include air temperature at $2$m, accumulated total evaporation, planetary boundary layer height, surface pressure, precipitable water for the entire atmosphere, UV-wind at $10$m and visibility. Land-use data include: elevation (derived from NASA's Shuttle Radar Topography Mission), water and forest land cover categories (derived from the National Land Cover Database 2011 as the percentage of each $1$km grid cell), population density, traffic density, and  PM\textsubscript{2.5} point and area-source emissions (obtained through the 2005 U.S. EPA National Emissions Inventory (NEI) facility emissions report \citep{epa2010}).


\subsection{Statistical Methods}
\subsubsection{The Linear Mixed Model (LMM)}
The PM\textsubscript{2.5}-to-AOD relation is known to vary in space and time, especially in a large area such as the northeast USA \citep{kloog2012incorporating}. For this reason, we follow current conventions \citep{kloog2012incorporating,kloog2014new} and introduce random space-time AOD-effects in our LMM. Recent studies \citep{kloog2015estimating, stafoggia2017estimation, de2018modelling} used LMM with random intercept and AOD-slope that vary in days and regions. Put differently, observations from a specific day-region combination have their own distribution, which differs from other regions in that day, or other days in that region. The formulation of this LMM implies independence of observations within day, and spatial independence between regions.

Our LMM model can be written as follows: 
\begin{equation} \label{eq:lmm}
	\begin{split}
    	& PM_{st} = \alpha + \sum_{k=1}^p \beta_k x_{k,st}  + (u_t +g_{rt}) +
    	(\beta_{AOD} + v_t + h_{rt}) AOD_{st}  +  \varepsilon_{st}, \\
    	& s \in \mathcal{S}, \ \ t \in (1,...,T), \ \ r \in (1,...,R),
    	\end{split}
\end{equation}
	
where $PM_{st}$ is the PM level at location $s$ at time $t$, $r$ encoding the region of location $s$, $\alpha$ a global average, the vector $(x_{k,st})_{k=1}^p$ encoding the attributes of location $s$ at time $t$, with corresponding effects $(\beta_k)_{k=1}^p$, and $u_t$ and $g_{rt}$ are day and day-region random effects. Finally, $AOD_{st}$ is the observed AOD in location $s$ and time $t$, with space-time varying random effect ($\beta_{AOD}+v_t+h_{rt}$), and $\varepsilon_{sj}$ is an independent and normally distributed error term. As it is very common in the literature, we assume here that $(u_t,v_t) \sim  N[0,\Sigma]$ and $(g_{rt},h_{rt}) \sim N[0,\Sigma_{REG}]$, where $\Sigma$ and $\Sigma_{REG}$ are diagonal $2 \times 2$ matrices.

Let us call $\eta^{\mbox{\tiny\itshape LMM}}_{st} = u_t + g_{rt} + (v_t + h_{rt}) AOD_{st} + \varepsilon_{st}$
the sum of random effects and the independent error for location $s$ at time $t$. $\eta^{\mbox{\tiny\itshape LMM}}$ is thus a vector of all the $N$ spatio-temporal random effects. We note that the LMM's definition imply a block-diagonal structure of $Var[\eta^{\mbox{\tiny\itshape LMM}}]$. This means that the covariance is fixed within regions, and independence between regions, as stated earlier.

Since PM assessment is done in retrospect, we do not intend to provide temporal forecasting, so our PM assessment task is essentially a prediction problem in the spatial domain of the study $\mathcal{S}$. Formally, we are interested in predicting PM value for every $t \in (1,...,T)$ and $s^* \in \mathcal{S}$. We emphasize that $s$ denotes the location of monitoring stations, i.e., locations seen in the data, whereas $s^*$ denotes locations previously unseen in our data.


\subsubsection{The Gaussian Markov Random Field (GMRF)}
As previously mentioned, GRFs may be infeasible to fit with large datasets. 
We thus opt for a GMRF approximation of the GRF. GMRFs are essentially multivariate Gaussian distributions, with a Markovian covariance, sampled at locations optimized to approximate a continuous GRF. The Markov property induce conditional independence between the random variables, so that the distribution at some point in a GMRF depends only on its set of neighbors. Conditional independence does not imply sparse covariance matrices, but it does imply a \emph{sparse precision matrix}, which is crucial to speed up computations \citep{rue2005gaussian}.

The Mat\'ern class of covariance functions is popular in the spatial statistics literature as it is a general class of stationary isotropic GRFs. A Mat\'ern covariance depends on the distances between points and is determined by parameters associated with its range and smoothness \citep{rasmussen2006gaussian}. \citet{lindgren2011explicit} show that the GMRF that approximates a GRF with a Mat\'ernan covariance is the solution of a particular set of \textit{stochastic partial differential equations} (SPDEs).

\begin{figure}[ht!]
\centering
  \includegraphics[width=110mm]{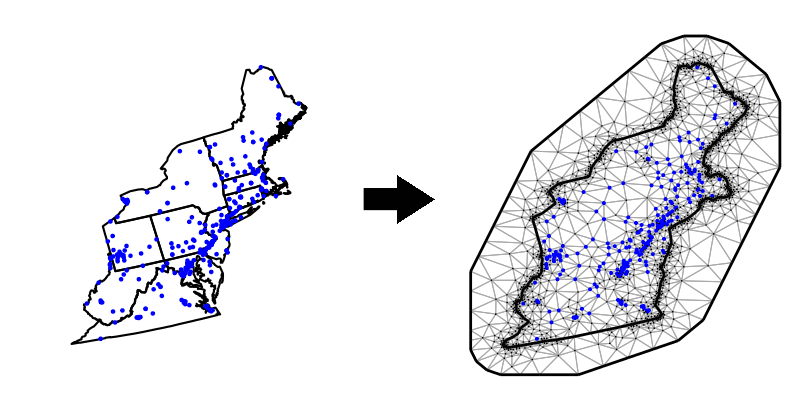}
  \caption{\footnotesize Study spatial domain (left) is replaced by FEM mesh consisted of triangles (right). Blue points indicate monitoring stations' location. To enable appropriate boundary conditions, the spatial domain is slightly expanded.}
  \label{fig:map2mesh}
\end{figure}

In a GMRF the continuous domain of the GRF, $\mathcal{S}$, is replaced by a discrete set of non-intersecting triangles using the Finite Element Method \citep{brenner2007mathematical}, as we illustrate in Figure~\ref{fig:map2mesh}. This representation of the spatial domain (along with some appropriate boundary conditions), together with the SPDE approach, allows fitting a GRF within the LGM framework, while enjoying the computational properties of a GMRF representation. 

GMRF approximations of GRF are gaining popularity in spatial statistics. The \textsf{R} package \textsf{R-INLA} \citep{rue2014inla,rue2017bayesian} has an efficient implementation of this approach using \textit{integrated nested Laplace approximations} (INLA) for fitting, and is particularly suitable for spatio-temporal models. We employ \textsf{R-INLA} along with \textsf{PARDISO} \citep{petra2014augmented} package for efficient sparse computations, in our satellite-based PM assessment.

In this paper our focus is on the statistical performance of LMM versus GMRF. Our GMRF replaces LMM's region-wise random effect with a Mat\'ern random field. The Mat\'ern field allows estimating for each day $t$, a spatial random intercept and AOD-slope at every location $s^* \in \mathcal{S}$, that does not depends on $s^*$'s region, $r$. For a fair comparison, we include a day-specific random intercept ($u_t$) and AOD-slope ($v_t$), assuming temporal independence as in the LMM. Also, we adhere to the random effects' intercept and slope independence assumption expressed in $\Sigma$ and $\Sigma_{REG}$ diagonal structures.

The only difference in terms of our formulation of the LMM and GMRF lies in the form of the spatial random effects. On one hand, an LMM with region-wise discrete random effects, and on the other hand, a GMRF with a continuously, smooth spatial field derived by a Mat\'ern covariance function.

Following the above notations, our GMRF can be described as:
\begin{equation} \label{eq:lgm}
	\begin{split}
      & PM_{st} = \alpha + \sum_{k=1}^p \beta_k x_{k,st} + (u_t + \gamma_{st}) +
      (\beta_{AOD} + v_t + \psi_{st})AOD_{st} + \varepsilon_{st}, \\
      & s \in \mathcal{S}, \ \ t \in 1,...,T.
    \end{split}
\end{equation}
where $u_t$ and $v_t$ are zero mean day-specific random intercept and AOD-slope assumed independent as in our LMM: $(u_t,v_t) \sim N[0,\Sigma]$ with diagonal $\Sigma$. $\gamma_{st}$ and $\psi_{st}$ are spatial latent fields assumed to be temporally independent i.e.:

\begin{equation}
   { Cov(\gamma_{st},\gamma_{s't'}) = \begin{cases} 0 & t \ne t'\\ 
          C(d_{ss'};\theta_{\gamma}) & t=t' \end{cases}, \\  \quad 
      Cov(\psi_{st},\psi_{s't'}) = \begin{cases} 0 & t \ne t'\\
          C(d_{ss'};\theta_{\psi}) & t=t' \end{cases}},
\end{equation}
where $C$ is the Mat\'ern covariance function, $d_{ss'}$ is the Euclidean distance between locations $s$ and $s'$, and $\theta_{\gamma}$ and $\theta_{\psi}$ are the hyperparameters of the Mat\'erns' correlation, governing range and scale. 

Like above, we define $\eta^{\mbox{\tiny\itshape GMRF}}_{st} = u_t + \gamma_{st} + (v_t + \psi_{st}) AOD_{st} + \varepsilon_{st}$, and $\eta^{\mbox{\tiny\itshape GMRF}}$ the vector of all $N$ $\eta^{\mbox{\tiny\itshape GMRF}}_{st}$ elements. The difference between GMRF and LMM is thus well understood in covariance terms: $Var[\eta^{\mbox{\tiny\itshape GMRF}}]$ replace the block-diagonal structure of $Var[\eta^{\mbox{\tiny\itshape LMM}}]$ with covariances that decay like the Mat\'ern function of the Euclidean distance. In other words, $Var[\eta^{\mbox{\tiny\itshape GMRF}}]$ is not determined by discrete definitions of spatial regions -- it decays smoothly and is not ``slab'' structured.


\subsubsection{Model Validation}
Most model validation techniques, such as cross-validation (CV), require independence between train and test sets \citep{arlot2010survey}. Ignoring dependencies will result in over-optimistic error estimates, thus favoring  more complex models \citep{hawkins2004problem}. In satellite-based PM assessment, spatial dependencies are to be expected, so that CV may return biased error estimates. For this reason, we compared our GMRF to LMM under various CV schemes, to verify that conclusions are not specific to the cross-validation scheme.

Our first CV folding scheme is \textit{K-fold} with $K=10$, which is a common convention in the PM assessment literature. In K-fold CV the entire dataset is randomly divided into $K$ folds. For each fold, we fit a model on a $K-1/K$ portion of the data, and compute the prediction error on the remaining $1/K$.

In addition K-fold, we consider a more conservative folding approach which we call \textit{leave-p-out--h-block} (LPO-h-block). This approach follows \citet{burman1994cross}'s \textit{h-block} CV, originally proposed for temporally correlated data. Our adaptation from temporal to the spatial domain is quite natural and similar to the folding scheme proposed by \citet{telford2009evaluation}. In \textit{LPO-h-block}, for every CV iteration, a sample of $p$ test stations is chosen at random. The rest of the stations serve as a train set, except stations in a radius of $h$ from each of the $p$ test stations. This alleviates the (spatial) dependence between train and test sets. We chose $p$ and $h$ such that after data omission, the training-test ratio is 9:1. In other words, with \textit{LPO-h-block}, instead of randomly splitting the data into two sets, we determine training and test sets by random selection of stations, but with a certain degree of spatial separation controlled by $h$.


\section{Results and Discussion}


\subsection{The Spatial Random Effect in a Prediction}
\begin{figure}[ht!]
\centering
  \includegraphics[width=110mm]{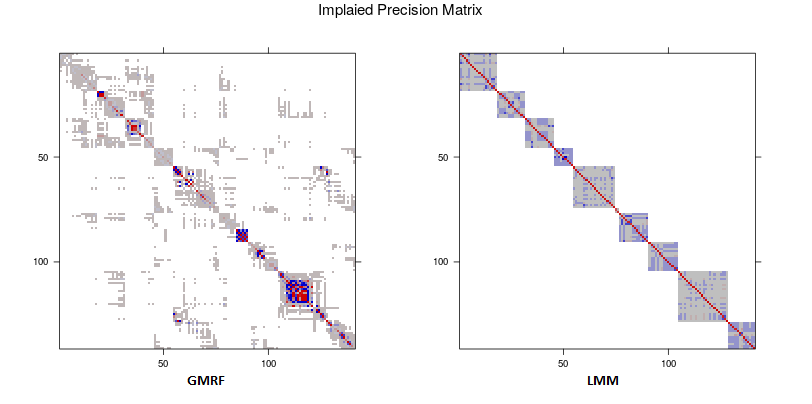}
  \caption{\footnotesize Part of the precision matrices of the GMRF (left) and LMM (right), describing the precision in one single day. The elements in the matrices are ordered by spatial regions. In contrast to LMM's region-wise spatial effect, the covariance between stations in the GMRF is not limited to a specific region and depends on spatial distance, this is expressed in a precision matrix that is not subject to discrete spatial definitions.}
  \label{fig:prec}
\end{figure}

Figure~\ref{fig:prec} is an illustration, in some arbitrary day, of the precision matrices implied by the two models. i.e., $(Var[\eta^{\mbox{\tiny\itshape GMRF}}])^{-1}$ on the left, versus  $(Var[\eta^{\mbox{\tiny\itshape LMM}}])^{-1}$ on the right. Matrices' elements are ordered according to the spatial region, so that the difference readily apparent. Both matrices are characterized by a similar level of sparsity, yet, the structure of GMRF's precision indicates that unlike in the LMM, some stations within the same region are uncorrelated, and some stations from different regions are correlated.

\begin{figure}[ht!]
\centering
  \includegraphics[width=110mm]{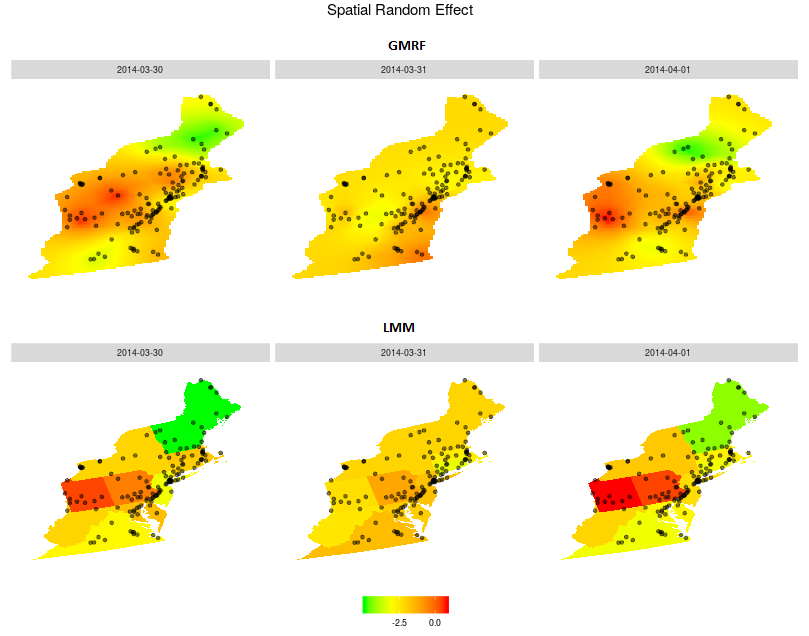}
  \caption{\footnotesize Spatial random effects distributions over the study domain, implied by GMRF (top) and LMM (bottom) for various days (columns). Color scale indicate the estimated value of the random effect. In LMM these values are confined to spatial units' defined regions, while in the GMRF they are free to vary across space.}
  \label{fig:re}
\end{figure}

Figure~\ref{fig:re} demonstrates the part in the PM's prediction due to spatial random effects in our data, i.e. the distributions of $g_{rt}$ and $\gamma_{st}$ over the study domain. The figure clearly demonstrates that LMM's usage of region-wise random effects returns a ``slab'' structured surface, which seems inconsistent with the underlying geography (Fig~\ref{fig:re} bottom). Particularly troubling is the fact that nearby stations, located in adjacent regions, get substantially different spatial effects. The GMRF, on the other hand, fit spatial random effects through a continuous Mat\'ern field that is smooth in space (Fig~\ref{fig:re} top).

\subsection{Prediction Error}
We evaluate prediction error of LMM and GMRF using two common measures: the root mean squared error (RMSE), and the coefficient of determination ($R^2$), both evaluated on the same holdout data sets. As previously mentioned, the holdout data sets were sampled according to different CV schemes: \emph{10-fold}, and \emph{LPO-h-block} with varying $h$ and $p$. Results are collected in Table~\ref{tab:CVres}. It can be seen that in our data, GMRF dominates LMM for all performance measures, and for all CV schemes. In particular using \textit{10-fold} CV, which is arguably the most prevalent in the PM assessment literature, the RMSE of GMRF is 10\% lower than that of LMM. The \textit{10-fold} $R^2$ of GMRF is 3 percentage points higher.

\begin{table}[ht]
\caption{PM\textsubscript{2.5} Prediction Accuracy: RMSE and R\textsuperscript{2} for different Cross Validation approaches}
\begin{center}
\begin{tabular}{cccccccccc}
    \hline
    && \multicolumn{2}{c}{\textit{10-fold} CV} && 
       \multicolumn{2}{c}{\textit{LPO-h-block} CV} && 
       \multicolumn{2}{c}{\textit{LPO-h-block} CV}  \\
    && && & \multicolumn{2}{c}{\textit{p=25, h=25\textsubscript{km}}} && \multicolumn{2}{c}{\textit{p=20, h=50\textsubscript{km}}} \\
    Model && RMSE & R\textsuperscript{2} && RMSE & R\textsuperscript{2} && RMSE & R\textsuperscript{2} \\
    \hline
    LMM && 2.68 & 0.80 && 3.12 & 0.74 && 3.13 & 0.73 \\
    GMRF && 2.42 & 0.83 && 2.79 & 0.79 && 2.86 & 0.77 \\
    \hline
\end{tabular} \\
\end{center}
\label{tab:CVres}
\end{table}

\begin{figure}[ht!]
\centering
  \includegraphics[width=110mm]{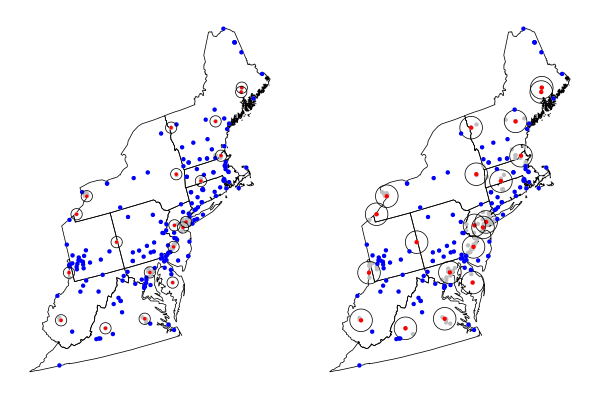}
  \caption{\footnotesize Example for one CV iteration in LPO-h-block folding with $h$=$25_{km}$ (left) and $h$=$50_{km}$ (right). Blue and red dots indicate stations included in the training and test sets, respectively. Grey dots indicate stations that were dropped.}
  \label{fig:lpohblock_map}
\end{figure}

The \textit{LPO-h-block} folding does not allow data measured at nearby geographic units to be found both in training and test set, as demonstrated in  Figure \ref{fig:lpohblock_map}. This folding scheme makes it more difficult to learn patterns that are related to a very specific spatial area, such as stations' nearest neighborhood, and which do not generalize to the entire spatial region. It appears that with \textit{LPO-h-block} CV, both GMRF's and LMM's performance decrease compared to \emph{10-fold} CV. The dominance of GMRF over LMM, however, remains. When the radius $h$ is 25\textsubscript{km}, GMRF's RMSE is still 10\% lower than LMM's, and the $R^2$ is still 5 percentage points higher. When we raise $h$ to 50\textsubscript{km}, i.e., when predictions are further apart in space, performance deteriorates, while conserving the dominance of GMRF. Although spatial dependence at this range is weaker, so that spatial random effects are smaller in magnitude, GMRF is still better (9\% lower than LMM's RMSE and 4 $R^2$ percentage points higher). 

The (statistically significant) finding that both models perform worse when evaluated using \textit{LPO-h-block} may suggest that a simple \textit{10-fold} returns overly optimistic error estimates. In other words, the success in predicting the test set was partly due to some extent of spatial overfitting. In order to examine the sensitivity of the results to the spatial dependence between the training and the test set, we consider different data splitting combinations, defined by the minimum distance between stations in training and test sets, by changing the value of $h$ in a \textit{LPO-h-block} CV. Clearly, when the spatial distance separating the training area from the testing area increases, the chance for overfitting decreases, but at the same time, the model is evaluated based on its ability to perform extrapolation more than interpolation. In extrapolation tasks, learning is aimed at achieving accurate predictions in more remote areas, yet usually at the expense of predictive power in nearby areas, for instance, through favoring models with less-complex and less-restrictive spatial dependence structures. Therefore, $h$ can be seen as governing the overfitting-extrapolation trade-off.

\begin{figure}[ht!]
\centering
  \includegraphics[width=90mm]{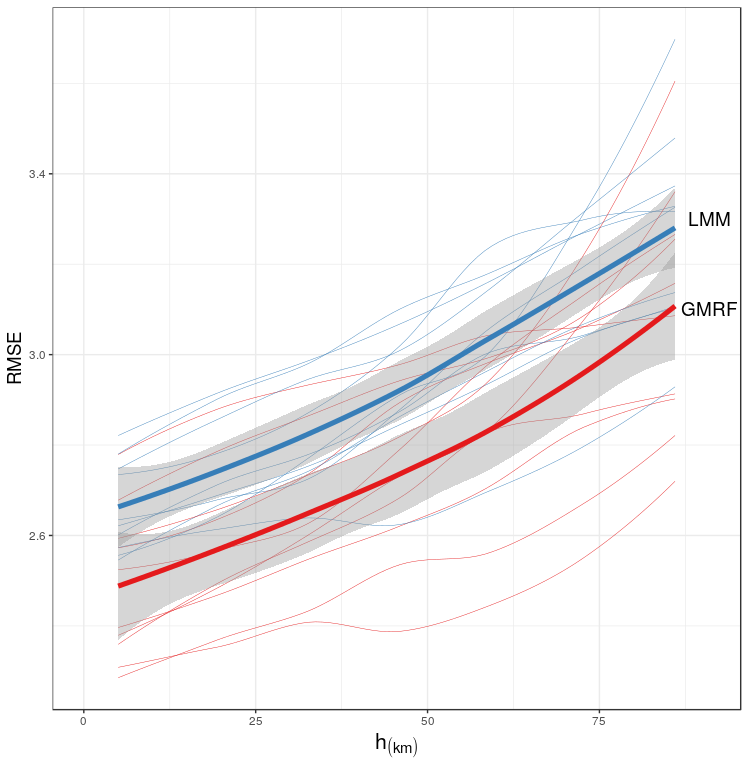}
  \caption{\footnotesize Prediction's RMSE is plotted against \textit{h-block}'s radius (km) in a LPO-h-block CV with $p = 20$ stations, over a data subset of the years 2010--2015. Thin lines are CV iterations, thick lines are average values and gray band is $95\%$ confidence interval.}
  \label{fig:CV_hblock}
\end{figure}

Is it possible that one model is better for short distance predictions, while the other is better for long distances (extrapolation)? 
Figure~\ref{fig:CV_hblock} shows that at any $h$, i.e., at any distance, GMRF dominates LMM. Put differently, GMRF is preferred if the research goal is accurate predictions in remote areas, or accuracy in areas where stations are crowded.

\section{Conclusions and future work}

\subsection{Conclusions}
In this paper we compared between an LMM with region-wise random effects, and a GMRF that approximates a GRF with Mat\'ern random field, for satellite-based PM prediction. We used data from the northeastern USA for the years 2000-2015 as a case study. The main difference between those models is expressed in the form of modeling spatial dependence. In LMM the spatial dependence is expressed through discrete random effects that are related to geographic defined areas as regions. In the GMRF, dependence is expressed through a Mat\'ern covariance function, allowing smoothly, continuously varied field of spatial random effects.

We have shown that in our data, the GMRF outperforms LMM. We find that this result is statistically significant for both RMSE and $R^2$ performance measures, and for various CV folding schemes. Specifically, we compared the popular \textit{10-fold} CV to a spatial-based leave-p-out CV, where train and test sets are spatially separated. Our analysis suggests that when the CV is more conservative, i.e., when training and test sets are less dependent, GMRF remains preferable. This latter finding suggests that including a Mat\'ern random field within a GMRF instead of discrete region effects in an LMM model, might be a better choice for predicting PM in both densely sampled and remote areas.

\subsection{Future Work}
While the dominance of GMRF over LMM was clear in our dataset, more datasets should be examined before generalizing our recommendations of GMRF. We showed that our findings are robust to the CV scheme. Choosing the appropriate CV, such that independence between train and test is ensured, is still a matter of active research.


\section*{Acknowledgements}

IK was supported by a research grant number 3-13142, from of the Ministry of Science and Technology, Israel.
JDR and RS were supported by grants 924/16 and 900/16 from the Israel Science Foundation.

\medskip

\section*{References}

\bibliographystyle{unsrt}
\bibliography{sample.bib}

\end{document}